\documentstyle[12pt,aaspp4,flushrt]{article}

\newcommand{\eg}{\mbox{e.g.}}
\newcommand{\kms}{\mbox{ km~s$^{-1}$}}

\def\gtorder{\mathrel{\raise.3ex\hbox{$>$}\mkern-14mu
             \lower0.6ex\hbox{$\sim$}}}
\def\ltorder{\mathrel{\raise.3ex\hbox{$<$}\mkern-14mu
             \lower0.6ex\hbox{$\sim$}}}
\def\arcs{\ifmmode {^{\scriptscriptstyle\prime\prime}}
          \else $^{\scriptscriptstyle\prime\prime}$\fi}
\def\arcm{\ifmmode {^{\scriptscriptstyle\prime}}
          \else $^{\scriptscriptstyle\prime}$\fi}
\newdimen\sa  \newdimen\sb
\def\parcs{\sa=.07em \sb=.03em
     \ifmmode $\rlap{.}$^{\scriptscriptstyle\prime\kern -\sb\prime}$\kern -\sa$
     \else \rlap{.}$^{\scriptscriptstyle\prime\kern -\sb\prime}$\kern -\sa\fi}
\def\parcm{\sa=.08em \sb=.03em
     \ifmmode $\rlap{.}\kern\sa$^{\scriptscriptstyle\prime}$\kern-\sb$
     \else \rlap{.}\kern\sa$^{\scriptscriptstyle\prime}$\kern-\sb\fi}
 
\begin{document}
 
\title{Why Quasar Pairs Are Binary Quasars \\ And Not Gravitational Lenses}
 
\author{ C.S. Kochanek, E.E. Falco  \& J.A. Mu\~noz}
\affil{Harvard-Smithsonian Center for Astrophysics \\
60 Garden Street \\ Cambridge, MA 02138, USA}
\authoremail{ckochanek@cfa.harvard.edu}
\authoremail{falco@cfa.harvard.edu}
\authoremail{jlehar@cfa.harvard.edu}
 
\begin{abstract}
We use simple comparisons of the optical and radio properties of the wide
separation ($3\arcsec < \Delta \theta < 10 \arcsec$) quasar pairs to demonstrate
that they are binary quasars rather than gravitational lenses.  The most
likely model is that all the pairs are binary quasars, with a one-sided 
2--$\sigma$ (1--$\sigma$) upper limit of 22\% (8\%) on the lens fraction.
Simple models for the expected enhancement of quasar activity during galaxy 
mergers that are consistent with the enhancement observed at low redshift can 
explain the incidence, separations, redshifts, velocity differences, and 
radio properties of the binary quasar population.  Only a modest fraction 
($\ltorder 5$\%) of all quasar activity need be associated with galaxy mergers
to explain the binary quasars. 
\end{abstract}


\section{Introduction}

Of the $\sim10^4$ known quasars (see, \eg, Hewitt \& Burbidge 1993), we know
of only $\sim40$ quasar pairs or multiples with separations smaller than
$10''$, most of which are confirmed gravitational lenses
(see Keeton \& Kochanek 1996).\footnote{A current summary of the
lens data is available at \tt{http://cfa-www.harvard.edu/glensdata}.} 
Almost all of the confirmed lens systems have separations smaller than $3''$ 
and are produced by relatively isolated, normal galaxies (see Keeton, 
Kochanek \& Falco 1997). Only 3 clear cases of multiple imaging by groups 
or clusters are known (Q~0957+561, Walsh et al. 1979; HE~1104--1805, Wisotzki 
et al. 1993; and MG~2016+112, Lawrence et al. 1984) and in all 3 cases there 
is at least one normal galaxy associated with the primary lens.  The remaining objects 
are wide-separation quasar pairs 
($3\arcsec < \Delta \theta < 10 \arcsec$), with similar (but not identical) 
optical spectra, and small velocity differences ($|\Delta v| \ltorder 10^3\kms$)
that lack a normal group or cluster of galaxies to act as the lens.  
In the Large Bright Quasar Survey (LBQS) there are 2 quasar pairs in a
sample of $10^3$ quasars, so the probability of finding an optically-selected 
quasar pair is $P_{pair} \sim 2 \times 10^{-3}$ (see Hewett et al. 1997).  
We get about the same value if we note that 11 pairs have been found 
in a total sample of $\sim 10^4$ quasars, but the estimate from the LBQS has 
the advantage of uniform selection criteria.  The pair fraction is 
two orders of magnitude higher than naively predicted from 
the quasar-quasar correlation function on Mpc scales (see Djorgovski 1991).

These problematic pairs have wreaked havoc on any quantitative discussion of 
gravitational lensing by group and cluster mass objects for over a decade.  
Determining whether these systems can be lenses is important because they 
require a previously unknown class of dark mass concentrations as the lens, and 
because the incidence of wide separation lenses is closely related to the 
amplitude of the power spectrum on $8 h^{-1} $ Mpc scales (i.e. $\sigma_8$).  
Standard cosmological models normalized to COBE or the local cluster abundance 
predict few wide separation lenses (e.g. Narayan \& White 1988; Cen et al. 1994;
Wambsganss et al. 1995; Kochanek 1995; Tomita 1996; Flores \& Primack 1996; Maoz et al. 1997) 
and require that most of the wide separation quasar pairs be binary quasars rather than 
gravitational lenses.  The pairs also lead to bizarre results about the 
structure of lenses if they are simply treated as gravitational lenses
(e.g. Park \& Gott 1997; Williams 1997).   

We list the known wide-separation quasar pairs in Table 1, and Figure 1 shows their
distribution in separation and redshift.  We have classified the pairs
in two ways.  First, we have divided the sample into the definite lenses,
the definite binary quasars, and the ambiguous pairs.    We can be certain a
quasar pair is a lens if the optical and radio flux ratios are consistent, the
velocity difference between the quasars is consistent with zero, and we see
a plausible lens candidate.  With these criteria we find the three quasar
lenses with separations larger than $3\arcsec$ from the first paragraph:
Q~0957+561, MG~2016+112, and HE~1104--1805.  We can be certain a quasar 
pair is a binary quasar if there is no plausible lens candidate and the 
optical and radio flux ratios are grossly discrepant (PKS~1145--071, 
MGC~2214+3550, and Q~1343+2640), or if there is no plausible lens candidate 
and there is a significant emission line redshift difference confirmed by 
an absorption feature at the same velocity in the spectrum of the foreground 
object (Q~0151+048=PHL~1222=UM~144).  
The remaining 10 objects lie between these two
regions of certainty:  they lack a plausible lens galaxy, the optical and
radio flux ratios do not grossly conflict, the velocity differences
are small or depend only on emission line centroids, and the spectra
show various levels of differences in their continuum and emission line
structures.  We tried to be extremely conservative in assigning objects to the
binary quasar class, and we deliberately ignored strong evidence
that several other pairs are binary quasars (e.g. MG~0023+171 whose morphology
is inconsistent with lensing, and HS~1216+5032 in which only one of the 
quasars is a BAL quasar, Q~1120+0195 which has a significant velocity difference,
and Q~0151+048, Q~1120+0195, LBQS~1429--008, and LBQS~2153-2056 which have
highly unlikely flux ratios for gravitational lenses).  
Second, we can classify the pairs based on their optical
and radio properties: systems in which both quasars are radio-faint 
($O^2$ pairs), systems in which one quasar is radio-bright ($O^2R$ pairs), 
and systems in which both quasars are radio-bright ($O^2R^2$ pairs).  We
call a quasar radio-bright if it is detected in the radio at a given
flux limit, and radio-faint if it is undetected, rather than radio-loud
and radio-quiet (which are defined by the ratio of optical and radio fluxes).
Where radio data were not already available, we searched for the pairs in the 
FIRST (White et al. 1997) and NVSS (Condon et al. 1996) surveys at 20~cm 
and 21~cm.  We discovered that Q~1343+2640 
is in fact an $O^2R$ pair, but that the remaining $O^2$ pairs are all 
radio-faint to the NVSS catalog limits of $\sim3$ mJy. By definition, 
lenses can only be $O^2$ or $O^2R^2$ pairs, and all $O^2R$ pairs
must be binary quasars. 

We will not argue about whether any individual ambiguous pair is a gravitational 
lens or a binary quasar.  Although observations and debates on these pairs are 
worthwhile, it is fair to say that detailed optical examinations of individual pairs 
have so far failed to produce convincing evidence for either the lens or
the binary hypothesis once the possibility of ``dark lenses'' is 
accepted.\footnote{An illustration of the difficulty was our internal debate over
whether Q~1120+0195=UM~425 deserved a ``?--'' designation as a pair with
strong evidence suggesting it is a binary rather than a lens.  EEF and JAM 
opposed the designation (see Michalitsianos et al. 1997).}    
Instead, we show in \S2 that the gravitational lens 
hypothesis makes predictions about radio properties of the quasar pairs that 
disagree with the data, while the binary quasar hypothesis naturally reproduces 
the data.  In \S3 we outline the relationship between quasars and galaxy 
mergers needed to explain the prevalence of binary quasars, and in \S4 we discuss the
implications of our conclusions.

\section{Why Quasar Pairs Cannot Be Lenses And Must Be Binaries}

\begin{deluxetable}{lccccccccc}
\footnotesize
\tablewidth{0pt}
\tablecaption{Wide Separation Quasar Pairs }
\tablehead{ Name &$z_s$ &$\Delta \theta$ &$R$ ($h_{50}^{-1}$ kpc) &$\Delta m$ &$f_R$ &$|\Delta v|$ ($\kms$)  &Lens? &Type &Ref }
\startdata
\tableline
 MG~0023+171                &$0.95$ &$4\farcs8$ &$40$ & 1.2 &$\sim10$ &$292\pm260$    &?--  &$O^2R^2$ &1  \nl
 Q~0151+048$^{\bf\dag}$     &$1.91$ &$3\farcs3$ &$28$ & 3.6 &         &$520\pm160$    &No   &$O^2$    &2  \nl
 QJ~0240--343               &$1.41$ &$6\farcs1$ &$52$ & 0.8 &         &$250\pm180$    &?    &$O^2$    &3  \nl
 RXJ~0911.4+0551            &$2.80$ &$3\farcs1$ &$24$ & 0.7 &         &$158\pm1000$   &?    &$O^2$    &4  \nl
 Q 0957+561                 &$1.41$ &$6\farcs1$ &$52$ & 0.2 & $1.3$   &$200\pm15$     &Yes  &$O^2R^2$ &5  \nl
 HE~1104--1805              &$2.32$ &$3\farcs1$ &$24$ & 1.7 &         &$300\pm90$     &Yes  &$O^2$    &6  \nl
 Q~1120+0195$^{\bf\dag\dag}$&$1.46$ &$6\farcs5$ &$56$ & 5.6 &         &$628\pm120$    &?--  &$O^2$    &7  \nl
 PKS~1145--071              &$1.35$ &$4\farcs2$ &$36$ & 0.8 &$>500$   &$200\pm110$    &No   &$O^2R$   &8  \nl
 HS~1216+5032               &$1.45$ &$9\farcs1$ &$78$ & 1.8 &         &$260\pm1000$   &?--  &$O^2$    &9  \nl
 Q~1343+2640$^*$            &$2.03$ &$9\farcs5$ &$78$ & 0.1 &$>57$    &$120\pm890$    &No   &$O^2R$   &10 \nl
 LBQS~1429--008             &$2.08$ &$5\farcs1$ &$42$ & 3.1 &         &$260\pm300$    &?--  &$O^2$    &11 \nl
 Q~1635+267                 &$1.96$ &$3\farcs8$ &$32$ & 1.6 &         &$33\pm86$      &?    &$O^2$    &12 \nl
 MG~2016+112                &$3.27$ &$3\farcs6$ &$26$ & 0.6 &$\sim1$  &$40\pm100$     &Yes  &$O^2R^2$ &13 \nl 
 Q~2138--431                &$1.64$ &$4\farcs5$ &$38$ & 1.2 &         &$0\pm115$      &?    &$O^2$    &14 \nl
 LBQS~2153--2056            &$1.85$ &$7\farcs8$ &$64$ & 2.9 &         &$1100\pm1500$  &?--  &$O^2$    &15 \nl
 MGC~2214+3550              &$0.88$ &$3\farcs0$ &$26$ & 0.5 &$>42$    &$148\pm420$    &No   &$O^2R$   &16 \nl
 Q~2345+007                 &$2.15$ &$7\farcs3$ &$58$ & 1.5 &         &$476\pm500$    &?    &$O^2$    &17 \nl
\tableline
\enddata
\tablecomments{ $z_s$ is the source redshift, $\Delta \theta$ is the angular separation,  
$R$ is the projected separation at the source redshift for
$\Omega_0=1$ and $H_0=50 h_{50} \kms$ Mpc$^{-1}$, $\Delta m$ is the magnitude difference
of the images, $f_R$ is the radio flux ratio or its limit if at least one quasar is
radio-loud, and $|\Delta v|$ is the velocity difference between the quasars. 
The entries in the Lens? column are: ``Yes'' if a normal lens
(galaxy, group, or cluster) is seen in the correct position to produce the
observed system, there is no significant velocity difference, and the radio
and optical data are consistent with the lens hypothesis; ``No'' if we see
no lens and either the radio emission or the emission line velocity difference,
confirmed by an absorption line velocity difference, 
are inconsistent with the lens hypothesis; and, ``?'' if we see no lensing object but have no
objective criterion to decide whether or not the object is lensed.  If there
is some evidence that the system is actually a binary, we used the label ``?--''.  
Type denotes the optical/radio classification of the pair.  Note that MG~2016+112
is really a triple system, not a pair. \\ 
$^{\bf\dag}$Q~0151+048 is also named PHL~1222 and UM~144.\\
$^{\bf\dag\dag}$Q~1120+019 is also named UM~425. \\
$^{*}$ We discovered that the brighter quasar is an $8.6$ mJy source at 20~cm, 
and the FIRST survey detection limit at the location of the fainter quasar 
is $0.15$ mJy, leading to a limit on the radio flux ratio of 57:1 as compared 
to an optical flux ratio of 1:1, making Q~1343+2640 an $O^2R$ pair. \\ 
References: (1) Hewitt et al. 1987, (2) Meylan et al. 1990, (3) Tinney 1995,
(4) Bade et al. 1997 (5) Walsh et al. 1979, (6) Wisotzki et al. 1993, (7) Meylan \& Djorgovski 1989,
(8) Djorgovski et al. 1987, (9) Hagen et al. 1996, (10) Crampton et al. 1988, (11) Hewett et al. 1989,
(12) Djorgovski \& Spinrad 1984, (13) Lawrence et al. 1984, (14) Hawkins et al. 1997, (15) Hewett et al. 1997,
(16) Mu\~noz et al. 1997, (17) Weedman et al. 1982. }
\end{deluxetable}

There are two independent lines of argument that force us to conclude
that the wide-separation quasar pairs are binary quasars.  The first is the absence
of an $O^2R^2$ quasar pair population comparable to the $O^2$ population,
and the second is the existence of the three $O^2R$ binary quasars.
We first outline the two arguments, and then set a 
statistical upper limit on the fraction of the quasar pairs that can
be gravitational lenses.
 
Under the lens hypothesis, the absence of a population of $O^2R^2$ quasar pairs is very puzzling
because the radio lens surveys (e.g. Burke, Leh\`ar \& Conner 1992;
King \& Browne 1996; Browne et al. 1997) have in fact found the majority of
gravitational lenses.  If we inventory the gravitational lenses with 
image separations smaller than $3\arcsec$, we know of 30 gravitational 
lenses whose sources are quasars or radio sources (see the summary in 
Keeton \& Kochanek 1996).  Twelve of these lenses are radio-faint 
quasars, and 18 are radio sources of which at least 8 are radio-bright quasars.  
The ratio of the numbers of radio and optical lenses at small separations, $r_s=1.5$,
should be maintained at larger separations given the comparable
redshift distributions and assuming similar angular selection functions.
The radio lens surveys have fairly uniform 
selection functions out to $30\arcsec$, and should have higher completeness 
than the optical quasar surveys rather than the reverse. For example,  
given the one certain $O^2$ lens (HE~1104--1805) we would expect $r_s=1.5$ 
$O^2R^2$ lenses, while we found two (Q~0957+561 and MG~2016+112).
The scaling works well, with a ``Poisson likelihood'' of 25\%.
If we take the 9 ambiguous $O^2$ pairs and interpret them all as 
``dark'' gravitational lenses, we would expect to find $9r_s=13.5$
``dark'' radio lenses. We in fact found only one (dubious) candidate,
MG~0023+171, which has a Poisson likelihood $\sim 10^{-5}$. A nearly equivalent
calculation is to note that the radio lens surveys have examined 
approximately $10^4$ sources of which 25--50\% will be high redshift quasars.
If the number of radio-bright quasars is $N_{Rqso} \sim 5000$, then we should find 
$ P_{pair} N_{Rqso} \simeq 10$ ``dark'' $O^2R^2$ quasar lenses.
Again, we find only the one candidate, so the Poisson likelihood is
$\sim 5 \times 10^{-4}$.  Here we counted only lensed radio quasars rather
than all radio lenses, leading to a modestly weaker limit.  
Since the completeness of the radio surveys is 
greater than that of the optical surveys, we have understated the case 
against the lens interpretation.

The key difference between the gravitational lens and binary hypotheses
is that in the binary hypothesis we must introduce the low probability
that a quasar is radio-bright into the calculation -- only $P_{R30} \simeq 5\%$ 
($P_{R1}\simeq 10\%$) of quasars are radio sources at 3.6~cm radio fluxes above 30 (1) 
mJy  (Hooper et al. 1996; Bischof \& Becker 1997).  If we start from
a sample of 11 optically selected quasar pairs, we should find that only
$11 P_{R1}^2 = 0.1$ of them are $O^2R^2$ pairs at a flux limit $\sim 1$ mJy, consistent
with finding none to the NVSS flux limit of $2.5$ mJy at 21~cm.\footnote{For the
flat radio spectra typical of radio-bright quasars, a 20~cm flux of 2.5~mJy corresponds
to a 3.6~cm flux of 1--6~mJy. The fraction of radio-bright quasars varies slowly
with radio flux, so a lack of precision in the flux limits has little effect
on the estimates.} In the radio surveys, 
the expected number of binary quasars in the sample is still  $ P_{pair} N_{Rqso} \simeq 10$,
but the radio surveys only discover the binaries in which both components are radio-bright 
at 3.6~cm fluxes $\gtorder 30$ mJy.  Thus the expected number of $O^2R^2$ binaries,
$ P_{R30} P_{pair} N_{Rqso} \simeq 0.5$, is smaller by a factor of $P_{R30}$ and is
consistent with the discovery of only one candidate (MG~0023+171).

\begin{figure}[ht]
{\epsfxsize=15cm \epsfbox{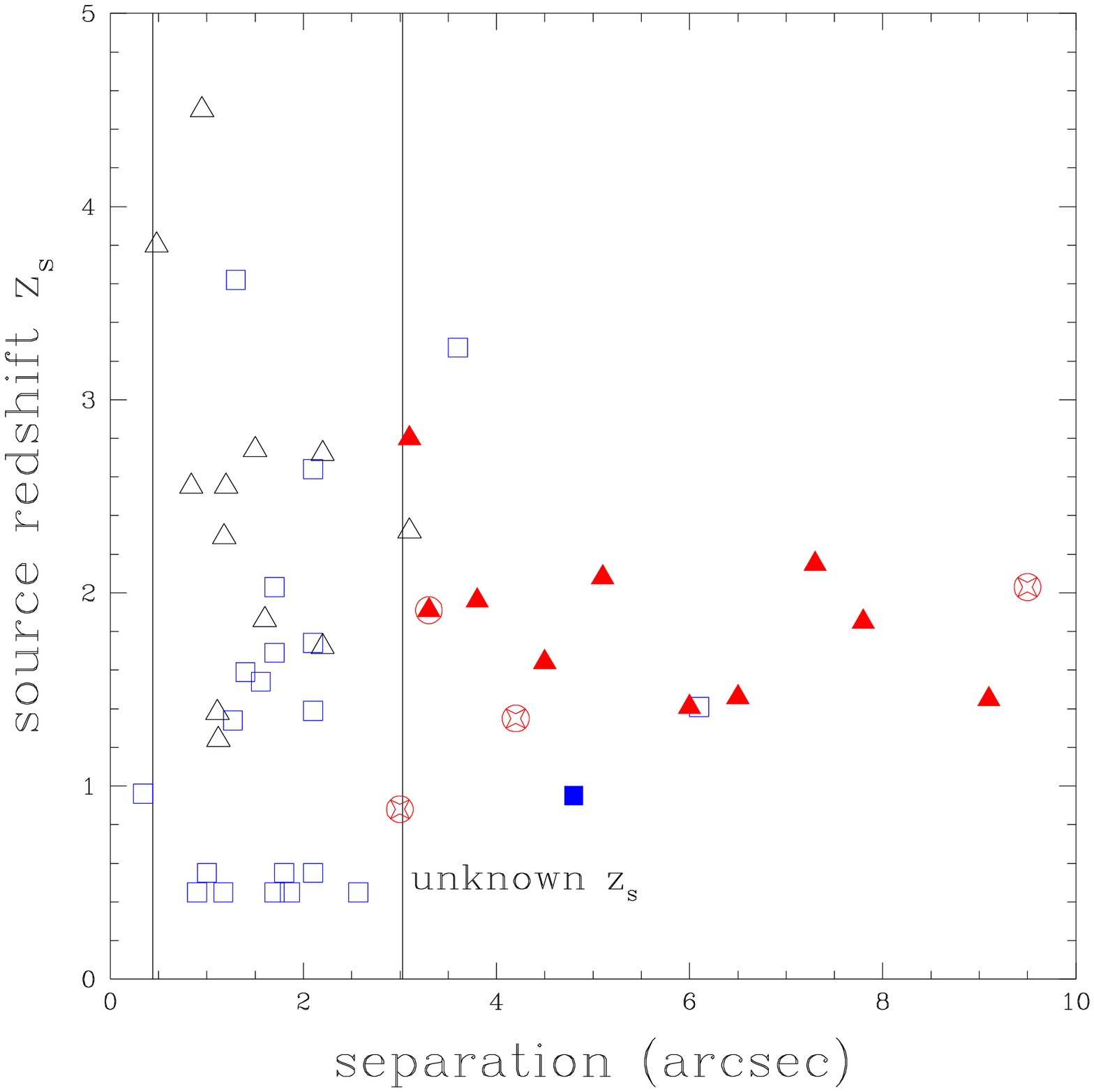}}
\figcaption{ The distribution of pairs, binaries and lenses in separation and source 
  redshift.  Triangles are radio-faint lenses and $O^2$ pairs, stars are $O^2R$
  binary quasars, and squares are radio-bright lenses and $O^2R^2$ pairs.  Open triangles
  and squares are certain lenses, filled triangles and squares are ambiguous
  pairs, and circled symbols are certain binaries.  If early-type galaxies 
  were the dominant lens population, 90\% of lenses would lie between the two vertical lines.  
  Radio lenses with unknown source redshifts are displayed in a band at $z_s=0.5$, although 
  we would expect their mean redshifts to be higher than for the lens systems with known 
  source redshifts.  }
\end{figure}

The existence of the $O^2R$ pairs is an equally important, independent argument
against the lens hypothesis, because there should be no $O^2R$ pairs in the
absence of a larger $O^2$ binary quasar population.  The relative numbers
of $O^2$, $O^2R$ and $O^2R^2$ binary quasars should be $(1-P_{R1})^2$, $2 P_{R1}(1-P_{R1})$ 
and $P_{R1}^2$.  Thus for 11 optically selected quasar pairs we would expect 
$2$ $O^2R$ pairs at a $\sim 1$~mJy flux limit, and we found one 
(Q~1343+2640), for a Poisson likelihood of 27\%. The radio surveys
should contain  $ (1-P_{R30}) P_{pair} N_{Rqso} \simeq 10$ $O^2R$ pairs,  
but they can only be found in optical follow up studies: PKS~1145--071 
was discovered in an optical search for multiple (lensed) images, and 
MGC~2214+3550 was discovered as part of a redshift survey of radio sources
which routinely took spectra of primary and secondary candidates for the
optical counterpart of the radio source.  The fraction of the radio
sources that have undergone some type of optical follow up study that
could detect an $O^2R$ pair is roughly $\epsilon_{opt}\sim 10\%$.  
We expect to find $\epsilon_{opt}(1-P_{R30}) P_{pair} N_{qso} \simeq 1$ 
$O^2R$ pairs in the existing data, so the discovery of two such pairs has 
a Poisson probability $\sim 18\%$.  Thus, the binary hypothesis naturally
explains the $O^2R$ pairs while they should not even exist under the
gravitational lens hypothesis.

So far, we have made the calculations assuming all pairs are either lenses
or binaries and found that the data strongly support the binary 
hypothesis.  We now combine these arguments to estimate the fraction
$f_L$ of quasar pairs that are gravitational lenses.  The optically
selected pair sample consists of $11 f_L$ lenses and $11 (1-f_L)$ 
binary quasars.  First, given the
11 optically selected pairs, we expect to find $11 f_L r_s = 16.5 f_L$ ``dark 
lenses'' in the radio surveys,  while with probability $f_L$ we found one and
with probability $1-f_L$ we found zero for the cases of treating 
MG~0023+171 as a lens or a binary.  Second, the binaries in the
optical sample should be divided as $ 11(1-f_L) (1-P_{R1})^2 = 10(1-f_L) $
$O^2$ binaries, $ 22 (1-f_L) P_{R1} (1-P_{R1}) = 2(1-f_L)$ $O^2R$ binaries,
and $ 11 (1-f_L) P_{R1}^2 = 0.1(1-f_L)$ $O^2R^2$ binaries, while
the optically-selected data consist of at least one $O^2$ binary, one $O^2R$ binary
and no $O^2R^2$ binaries for a flux limit $\sim 1 $ mJy.  Third, 
given $N_{Rqso} \sim 5000$ radio quasars we expect 
$ P_{R30} P_{pair} N_{Rqso} (1-f_L) \simeq 0.5 (1-f_L)$ $O^2R^2$ pairs for a flux 
limit $\sim 30 $ mJy, and we found either zero with probability $f_L$ or one 
with probability $1-f_L$ depending on the treatment of MG~0023+171.  We also expect 
$ \epsilon_{opt} (1-P_{R30}) P_{pair} N_{Rqso} (1-f_L) \simeq (1-f_L)$
$O^2R$ pairs and we found 2.  We then compute the likelihood distribution
for $f_L$ assuming a uniform prior.  The maximum likelihood value is
that all objects are binaries ($f_L=0$) with a one-sided 2--$\sigma$ 
(1--$\sigma$) Bayesian upper limit of $f_L<22\%$ (8\%).    

The result is robust even though our calculation includes a number of
crudely estimated parameters. For example, if we strongly bias 
the parameters to favor the lens hypothesis, by halving $r_s$ and 
doubling $N_{qso}$ and $\epsilon_{opt}$, the 2--$\sigma$ 
(1--$\sigma$) upper limits only shift to $f_L<40\%$ (14\%).  If we
drop the constraint from the expected number of ``dark'' radio lenses, 
and only determine the lens fraction consistent with the numbers of 
$O^2R$ pairs, we find $f_L < 69\%$ ($38\%$) at 2--$\sigma$ (1--$\sigma$).
In this model the maximum likelihood value for the lens fraction is
still $f_L=0$, but our limits are greatly weakened by our ultra-conservative 
division of the sample into binaries and ambiguous pairs.

There are several additional peculiarities to interpreting the $O^2$
pairs as a lensed population.  First, the two wide-separation radio 
lenses have normal objects as lenses.
In fact, all lenses imaged by HST besides the ambiguous $O^2$ pairs
show an obvious lens galaxy unless the quasar images are so bright as 
to prevent the detection of any reasonable galaxy due to the contrast 
(see the summary by Keeton et al. 1997).  Second, bright quasar
lenses should rarely show large flux ratios ($\gtorder 1$--$2$ mag 
depending on the magnitude, see Kochanek 1993) between the components, 
while four of the $O^2$ pairs show magnitude differences of more than 
2.5 mag (Q~1120+019, Q~1429--008, Q~0151+048, LBQS~2153--2056).  Third,
a significant fraction (30--40\%) of the small separation lenses are 
four-image rather than two-image lenses, while we find no four-image
wide-separation lenses.  The counterargument here is that the lack
of wide-separation four-image lenses is simply evidence that ``dark lenses''
have rounder potentials than galaxies.  Fourth, the probability for producing
lenses of a given separation rises monotonically with source redshift,
although the probability will rise faster at small separations since
the more massive lenses form at lower redshifts in standard models.  Thus
the concentration of all $O^2$ pairs near $z_s=1$ to $2$ is 
difficult to reconcile with the more uniform redshift distribution
of the smaller separation lenses.  The selection functions for finding
lenses and pairs are reasonably uniform inside $6\arcsec$ (the outer
radius checked by most optical lens surveys),  so the rise in the ratio of 
small separation quasar lenses to $O^2$ pairs from 1:1 to 4:1
between $z=2$ and $z=3$ (see Fig. 1) is strong evidence against the lens
hypothesis.

\section{Consequences of the Binary Hypothesis}

The standard criticism of the binary quasar explanation for the quasar pairs
was clearly stated by Djorgovski (1991).  For a comoving bright quasar density
of $n_q \simeq 500 h_{50}^3 $ Gpc$^{-3}$ (Hartwick \& Schade 1990) and a
correlation function $\xi(r) = (r/r_0)^{-1.8}$ with
$r_0 \simeq (11\pm2)h_{50}^{-1}$ Mpc (Croom \& Shanks 1996),  the probability of
finding a second quasar within projected separation $R=50 h_{50}^{-1} R_{50}$ 
kpc is $P_{q0} \sim 2 \times 10^{-5} R_{50}^{1.2}$.  The observed probability
of finding a close quasar pair of $P_{pair} \sim 2 \times 10^{-3}$ 
(e.g. Hewett et al. 1997) is a factor of $\sim 10^2$ larger.  The enhancement
is confined to very small spatial scales, because we should have found
20 times as many pairs between $10\arcsec$ and $100\arcsec$ as are found between
$3\arcsec$ and $10\arcsec$ if the distribution simply followed the slope 
of the correlation function.  Wide area surveys such as the LBQS have not found such a 
larger quasar pair population (given the 2 wide-separation quasar pairs in the LBQS, they
should have found 40 wider pairs!). Thus, the binary quasars correspond to 
merging galaxies with separations $\ltorder 50 h_{50}^{-1}$ kpc rather than 
chance superpositions, and the fundamental flaw in using the correlation 
function argument against the binary hypothesis (as already noted by Djorgovski 1991) 
is that it fails to include the increased probability that a black hole will be an 
active quasar during a galaxy merger.  In the local universe we see that nuclear 
activity (starbursts and AGN) is enhanced by mergers and interactions (e.g. Keel 1996; 
Bahcall et al. 1997).  Theoretically, 
simulations of merging galaxies (e.g. Barnes \& Hernquist 1996) demonstrate 
that interactions drive gas toward the central regions of the galaxy where it 
can be used to fuel a new burst of activity in a previously quiescent galaxy.  
Far from being surprised, we should have predicted that binary quasars would 
be significantly more abundant than predicted by the correlation function on 
large (Mpc) scales.  

Quasars are the active cores of galaxies, so to understand the clustering of quasars
we should start from the clustering of the underlying galaxies.  The quasar-quasar
correlation function on large (Mpc) scales should be identical to the
galaxy-galaxy correlation function; so for a comoving galaxy density $n_g$,
the comoving density of galaxy pairs with separation smaller than $r_a$ is 
\begin{equation}
   n_{g2} = 2\pi n_g^2 \int_0^{r_a} r^2 \xi(r) = 5.2 n_g^2 r_a^{1.2} r_0^{1.8}
\end{equation}
provided the probability for a third galaxy in the volume is small (e.g. Peebles
1993).  If the probability $f_{iso} \ll 1$ of an isolated galaxy having an active 
quasar is the same for all galaxies, then $n_q = f_{iso} n_g$ and
the density of quasar pairs is $n_{q2} = f_{iso}^2 n_{g2}$.  For a constant 
comoving galaxy density of $\sim 10^6 h_{50}^3$ Gpc$^{-3}$ (e.g. Marzke et al. 1994; 
Loveday 1992) we find $f_{iso} \sim 5 \times 10^{-4}$.  The fraction 
of quasar pairs, $n_{q2}/n_q \simeq 5 f_{iso} n_{g2}/n_g \simeq 5 n_q r_a^{1.2} r_0^{1.8}$,
with separations smaller than $r_a$ is consistent with our earlier, direct estimate from
the quasar-quasar correlation function.  Suppose, however,
that the probability of a galaxy having an active quasar when it is a member
of a pair is $f_{merge} > f_{iso}$.  The
density of quasar pairs is now $f_{merge}^2 n_{g2}$, and the fraction of
quasar pairs is larger by the factor $\beta^2=(f_{merge}/f_{iso})^2$.  Thus to explain
the $10^2$ overabundance of quasar pairs we need only a factor of $\beta=10$ increase
in the probability of quasar activity in merging systems over isolated systems.

At low redshift, it is known that the amplitude of the galaxy-quasar correlation function 
is larger than the amplitude of the galaxy-galaxy correlation function 
by a factor of $4\pm1$ (Fisher et al. 1997; Yee \& Green 1987; French \& Gunn 1983).  
Equivalently, the fraction of quasars with companion galaxies, $ \beta n_{g2}/ n_g$, 
is larger than the fraction of galaxies with companions, $n_{g2}/n_g$, by a factor of 
$\beta=f_{merge}/f_{iso}$.  The correlation function comparisons average the galaxy
density over volumes much larger than the physical scales on which tidal interactions
provide a mechanism for increasing the amount of quasar activity, so the ratio of the 
correlation function amplitudes sets only a lower bound of $\beta > 4$.  On small
scales we estimate enhancements of $\beta \sim 17$ ($9$) from the 2 companions
brighter than $L_*$ (or the 9 brighter than $0.1 L_*$) 
found within $50 h_{50}^{-1}$ kpc of the 20 bright, low-redshift quasars 
studied by Bahcall et al. (1997).  As expected, the enhancement
is larger than the limit derived from the correlation function measured
over larger volumes. The amplitude of the enhancement is sufficient to explain the incidence
of binary quasars in the high-redshift sample.  Note that we do not expect any
binary quasars in the Bahcall et al. (1997) sample because the expected number of 
binary quasars is smaller than the number of quasars with companion galaxies by
a factor of $f_{merge} \sim 5 \times 10^{-3}$ for $f_{iso} \sim 5 \times 10^{-4}$
and $\beta=10$.

Models of the history of quasars and supermassive black holes (Small \& Blandford 1992; 
Haehnelt \& Rees 1993) suggest that the enhancement in the number of binary quasars can be 
explained either by an increase in the formation rate of black holes, or by the reignition of
existing black holes as quasars.  The observed luminosity of 
the quasars is directly related to the final, total mass in supermassive black 
holes because the luminosity is set by the accretion rate, and the mean black hole 
mass per galaxy of approximately $10^7 M_\odot$ estimated from the total quasar
luminosity is consistent with the results of direct dynamical observations of nearby galaxies
(see Kormendy \& Richstone 1995; Magorrian et al. 1997).  At the Eddington limit,
a black hole's mass increases by a factor of 10 in 
$t_{10} \simeq 10^8 \epsilon_1$ yrs, 
where $\epsilon_1$ is the accretion efficiency in units of 10\%.  Since we do
not find $10^{10} M_\odot$ black holes locally, and the bright quasars making
up the pairs require $10^8$--$10^9 M_\odot$ black holes fueled near the
Eddington rate, $t_{10}$ sets a maximum active lifetime for the black
holes as bright quasars.  The age of the universe ($t=t_0 (1+z)^{-3/2}$ with 
$t_0 = (2/3)(c/H_0)$ for $\Omega=1$) is much longer than the lifetime 
of any individual quasar since 
$t_0/t_{10} \simeq 130 h_{50}^{-1} \epsilon_1^{-1} (1+z)^{-3/2}$.
Thus, the duty cycle, or probability that a massive black hole is functioning
as a luminous quasar is only a few percent.  At high redshift we can approximate
the black hole formation rate by a constant rate of 
$\dot{n}_{BH} \simeq n_Q/t_{10}$, and the total number of massive black holes 
by $n_{BH} \simeq n_Q (t_0/t_{10})(1+z)^{-3/2}$.  The quiescent black holes 
outnumber the quasars by a factor of $(t_0/t_{10})(1+z)^{-3/2}$.  

The simplest general model for the formation of quasars during mergers is
simply to add a density dependent term to the formation probability,
\begin{equation}
   f \simeq f_{iso} + f_{merge} \xi \left(\hbox{max}\left[r,r_a\right]\right).
\end{equation}
If we allow the second term to saturate at $r_a \ltorder 50 h_{50}^{-1}$ kpc
and use the low-redshift enhancement factor of $\beta=f_{merge}/f_{iso} \sim 10$,
mergers naturally explain the incidence of high redshift binary quasars. 
To the extent that formation and renewed accretion are qualitatively
different processes, an alternate process is to form the binary quasars by 
reigniting extinct quasars during mergers.  The probability of finding a black hole near
a quasar is larger than the probability of finding a quasar by the factor
$(n_{BH}/n_q) \gg 1$.  If $p_{ri}$ is the probability of reigniting an
extinct quasar, then $\beta = p_{ri} n_{BH}/n_q$, and 
\begin{equation}
  p_{ri} \simeq 10 { t_{10} \over t_0 } (1+z)^{3/2} \simeq 0.1(1+z)^{3/2}. 
\end{equation}
is sufficient to explain the numbers of binary quasars.  In either case,
only a few percent of all quasars are formed by the merger related processes, 
so they do not represent a significant change in the mean quasar formation history.

Unlike the gravitational lens hypothesis, the binary hypothesis
naturally explains the concentration of the pairs between
$1 \ltorder z \ltorder 2$.  In both formation models the peak in the binary
quasar redshift distribution should be near the peak in the quasar
redshift distribution, with a comparable or narrower width.  The
density dependent formation model produces a peak at the same
redshift, while the reignition model should have a peak at slightly
lower redshift because the number of extinct quasars is higher on
the low redshift side of the quasar peak. In the Boyle,
Shanks \& Peterson (1988) quasar luminosity function, the peak
surface density at $B\simeq 19$ mag is at $z\simeq 2$ and 90\% of
the quasars lie at $z \ltorder 2.5$, very similar to the observed
distribution of the pairs.  Both models predict that binary quasars 
should be significantly rarer than gravitational lenses at $z >2.5$ 
even though they are almost equally abundant at $z=2$.  The binary
hypothesis also provides a natural explanation for the wide
range of optical flux ratios seen in the quasar pairs.
While it is unlikely for a lens system with a bright quasar image
($m \ltorder 19$ B mag) to have a large
flux ratio, variations in the accretion rates and black hole masses 
provide a natural explanation for the broad range of flux ratios
seen in the quasar pairs.  If we drew the luminosities
of the pairs randomly based on the luminosity function we would expect
most binary quasars to show large flux ratios, constrained by
the dynamic range limits of the quasar surveys (see Kochanek 1995). 
The common triggering mechanism for the quasar binaries may, however,
produce luminosity correlations between the two quasars. 

Like the gravitational lens hypothesis, the
binary hypothesis naturally explains the separations of the pairs and the
small velocity differences.  The use of mergers to trigger more 
quasar activity requires characteristic separations $\ltorder 50 h_{50}^{-1}$ kpc.  
Although the pairs will tend to have larger separations than their
orbital pericenters because they spend more time near apocenter, we should
expect to find some binary quasars below $\Delta\theta=3\arcsec$.  The
apparently sharp cutoff at separations larger than $10\arcsec$ may
also be more characteristic of mergers than of gravitational
lenses -- statistical models of wide separation lenses (see Kochanek 1995;
Wambsganss et al. 1995) that produce the observed numbers of pairs 
as lenses generally have a slowly declining distribution in separation.  
The velocity differences between the quasars should be characteristic of binary
galaxies.  For example, the mean pair-wise velocity
dispersion in the CfA redshift survey is $295\pm100\kms$ if Abell 
clusters with $R > 1$ are excluded from the sample (Marzke et al. 1995),
and the typical difference for Seyferts in binary galaxy systems
is $170\pm200$ (Keel 1996), both of which are consistent with the
observed velocity differences of the pairs.

\section{Discussion}

Our comparison of the optical and radio data rules out the pure 
gravitational lens hypothesis for the quasar pair population,
and requires that most of the quasar pairs be binary quasars.  
However, it is a statistical argument about the population, and 
it cannot prove that any individual quasar pair is binary rather
than a lens.  On the other hand, once we demonstrate that most of the pairs
must be binary quasars, Occam's razor suggests that they are all
binary quasars, since the ``dark lens'' hypothesis now requires an
entirely new  population of objects with the masses of clusters
but no stellar or X-ray luminosity to produce a few wide separation
``dark lenses'' that are essentially indistinguishable from a 
dominant population of binary quasars.  The enormous consequences of 
even a modest population of ``dark lenses'' demands a high standard of 
proof before invoking the interpretation for any pair.  If there are 
$\simeq 3$  lenses produced by a ``dark lens'' population (scaling from 
Wambsganss et al. 1995; Kochanek 1995) the comoving density of the 
``dark lenses'' matches that of groups and clusters, the true value of 
$\sigma_8$ is significantly larger than estimated from the abundance 
of clusters, and many results in structure formation and about the shape 
and amplitude of the power spectrum become invalid.

The association of quasar activity with mergers is an old idea, as is the 
suggestion that the quasar pairs are related to merger activity (e.g. Djorgovski 
1991).  It is seen in the local universe where nuclear activity is more common in merging 
systems (e.g. Keel 1996; Bahcall et al. 1997), and it is expected from theoretical models 
of gas dynamics during mergers (e.g. Barnes \& Hernquist 1996).  In fact, the
enhancement in quasar activity produced by mergers estimated from the number
of galaxies seen near the Bahcall et al. (1997) quasar sample exactly matches
the enhancement needed to explain the binary quasar population at high redshift.
Furthermore, the merger model naturally explains all the features of the 
binary quasar population: (1) the separations are characteristic of scales on
which tidal perturbations become important ($R \ltorder 50 h_{50}^{-1}$ kpc);
(2) the radio properties of the sample are explained by
the low probability that quasars are radio-bright; (3) the relative
velocities should be $\ltorder 10^3 \kms$ because they are 
characteristic of merging galaxies; (4) the pairs should be
concentrated in redshift at or below the peak in the quasar abundance.

While we avoided arguments about the interpretation of the individual, ambiguous 
pairs in our discussion, it is still important to examine the individual
cases.  The only simple test that can unambiguously prove that a quasar pair
is a binary is to show that it is an $O^2R$ binary quasar using deep radio
observations. For example, Patnaik, Schneider \& Narayan (1996)  discovered that 
Q~2345+007~A is a $\sim 30$ $\mu$Jy radio source. The B image was not detected,
but the predicted flux was too close to the noise level to test the lens hypothesis.  
Nonetheless, similar observations of the other $O^2$ pairs 
should show that several of them are $O^2R$ pairs since the fraction of quasars with 
detectable radio flux begins to rise below a 3.6~cm flux of 1~mJy.  At 0.3~mJy the
fraction of detectable quasars is 15\% (Hooper et al. 1995), but little is
known about the fraction of quasars with emission at $\sim 0.01$ mJy 
(see Fomalont et al. 1991).
Measuring a time delay is the only unambiguous test to prove that a pair is 
a ``dark lens'', although a strong case can also be made if there is a significant
weak lensing detection centered on the pair.  A weak lensing effect is 
detected near Q~2345+007 (Bonnet et al. 1993), but its center is too far from 
the pair to be responsible for the image splitting.  The ambiguities
in current arguments about spectral similarities would be greatly reduced by  
quantitative spectral comparisons between random isolated quasars, binary quasar
members, lensed quasars and the ambiguous pairs.
An initial study by Small, Sargent \& Steidel (1997) showed that
the emission line differences in Q~1634+267 and Q~2345+007 were consistent with the
differences seen in the spectra of isolated quasars viewed at different times.
However, such comparisons must account for two biases.  First, in focusing on Q~1634+267
and Q~2345+007, Small et al. (1997) selected the quasar pairs already known to
show smaller spectral differences than a randomly selected quasar pair.  Second, a
proper comparison of the spectra of the various object classes must also compensate for
the possibility that binary quasars will be more similar than randomly selected
quasar pairs simply because they have similar redshifts, luminosities, and
environments.

\acknowledgements \noindent Acknowledgements: 
We thank J. Leh\`ar for reading the manuscript.
CSK is supported by NSF grant AST-9401722 and NASA ATP grant NAG5-4062.
Our research was supported by the Smithsonian Institution. 
JAM is supported by a postdoctoral fellowship from the Ministerio de
Educaci\'on y Cultura, Spain.

\end{document}